\documentclass[runningheads]{llncs}
\usepackage[T1]{fontenc}
\usepackage{graphicx}
\usepackage{gensymb} 

\begin{document}
\title{Empowering Mobility: Brain-Computer Interface for Enhancing Wheelchair Control for Individuals with Physical Disabilities}
\titlerunning{Empowering Mobility for Physical Impairment}
%
\author{Shiva Ghasemi\inst{1}\orcidID{0000-0003-1435-6342} \and
Denis Gra{\v{c}}anin\inst{1}\orcidID{0000-0001-6831-2818} \and
Mohammad Azab\inst{2}\orcidID{0000-0002-9386-4726}}

\authorrunning{S. Ghasemi et al.}
%
\institute{Virginia Tech, Blacksburg, VA 24060, USA 
\and
Virginia Military Institute, Lexington, VA 24459, USA
\\
\email{shivagh@vt.edu, gracanin@vt.edu, azabmm@vmi.edu}}

\maketitle              
\begin{abstract}
The integration of brain-computer interfaces (BCIs) into the realm of smart wheelchair (SW) technology signifies a notable leap forward in enhancing the mobility and autonomy of individuals with physical disabilities.
BCIs are a technology that enables direct communication between the brain and external devices.
While BCIs systems offer remarkable opportunities for enhancing human-computer interaction and providing mobility solutions for individuals with disabilities, they also raise significant concerns regarding security, safety, and privacy that have not been thoroughly addressed by researchers on a large scale.
Our research aims to enhance wheelchair control for individuals with physical disabilities by leveraging electroencephalography (EEG) signals for BCIs.
We introduce a non-invasive BCI system that utilizes a neuro-signal acquisition headset to capture EEG signals.
These signals are obtained from specific brain activities that individuals have been trained to produce, allowing for precise control of the wheelchair.
EEG-based BCIs are instrumental in capturing the brain's electrical activity and translating these signals into actionable commands.
The primary objective of our study is to demonstrate the system's capability to interpret EEG signals and decode specific thought patterns or mental commands issued by the user.
By doing so, it aims to convert these into accurate control commands for the wheelchair.
This process includes the recognition of navigational intentions, such as moving forward, backward, or executing turns, specifically tailored for wheelchair operation.
Through this innovative approach, we aim to create a seamless interface between the user's cognitive intentions and the wheelchair's movements, enhancing autonomy and mobility for individuals with physical disabilities.

\keywords{Accessibility \and BCI \and EEG \and Smart wheelchair}
\end{abstract}

\section{Introduction}
According to the Americans with Disabilities Act (ADA)~\cite{noauthor_ada_nodate}, a physical impairment encompasses any physiological disorder or condition, disfigurement, or anatomical loss impacting one or more of the body's systems.
This includes, but is not limited to, systems such as neurological, musculoskeletal, respiratory, reproductive, cardiovascular, or endocrine systems.
World Health Organization (WHO) states, approximately 15\% of the global populace experiences some form of disability~\cite{Maksud2017LowcostEB,WorldHealthOrganization_disability}.
Individuals with physical impairments may experience congenital disabilities or acquire such impairments later in life due to various etiologies, including accidents, diseases, the aging process, or military service.
These impairments, manifesting in forms such as limb loss, paralysis, or reduced mobility, can profoundly impact one's capacity to execute activities of daily living (ADLs) and instrumental activities of daily living (IADLs) – both of which are pivotal for autonomous living.
Extensive research, such as the studies by McGrath et al.~\cite{mcgrath2019impairments} and Lee et al.~\cite{lee2021impacts}, have delved into the impact of physical impairment on ADLs and IADLs.
These studies highlight the complex interplay between these functional domains and their consequential effects on an individual's overall functionality and well-being.

Aging often brings about various physical impairments.
Common age-related impairments include reduced mobility, muscle weakness, joint pain, and bone fragility.
Additionally, a decrease in spatial cognitive abilities has been associated with advancing age~\cite{klencklen2012we}.
There exists a link between physical and cognitive impairments.
Studies have demonstrated that greater physical activity is associated with lower incidence of cognitive impairment in later life.
For instance, research has indicated that individuals with physical disabilities, such as stroke survivors, may experience cognitive deficits, including memory impairment and executive dysfunction~\cite{lee2021impacts}.
In addition to mobility constraints, cognitive aging impacts the ability of older adults to find their way, affecting both their sense of direction and navigation skills~\cite{moffat2001age}.

\begin{figure}
\includegraphics[width= 5cm]{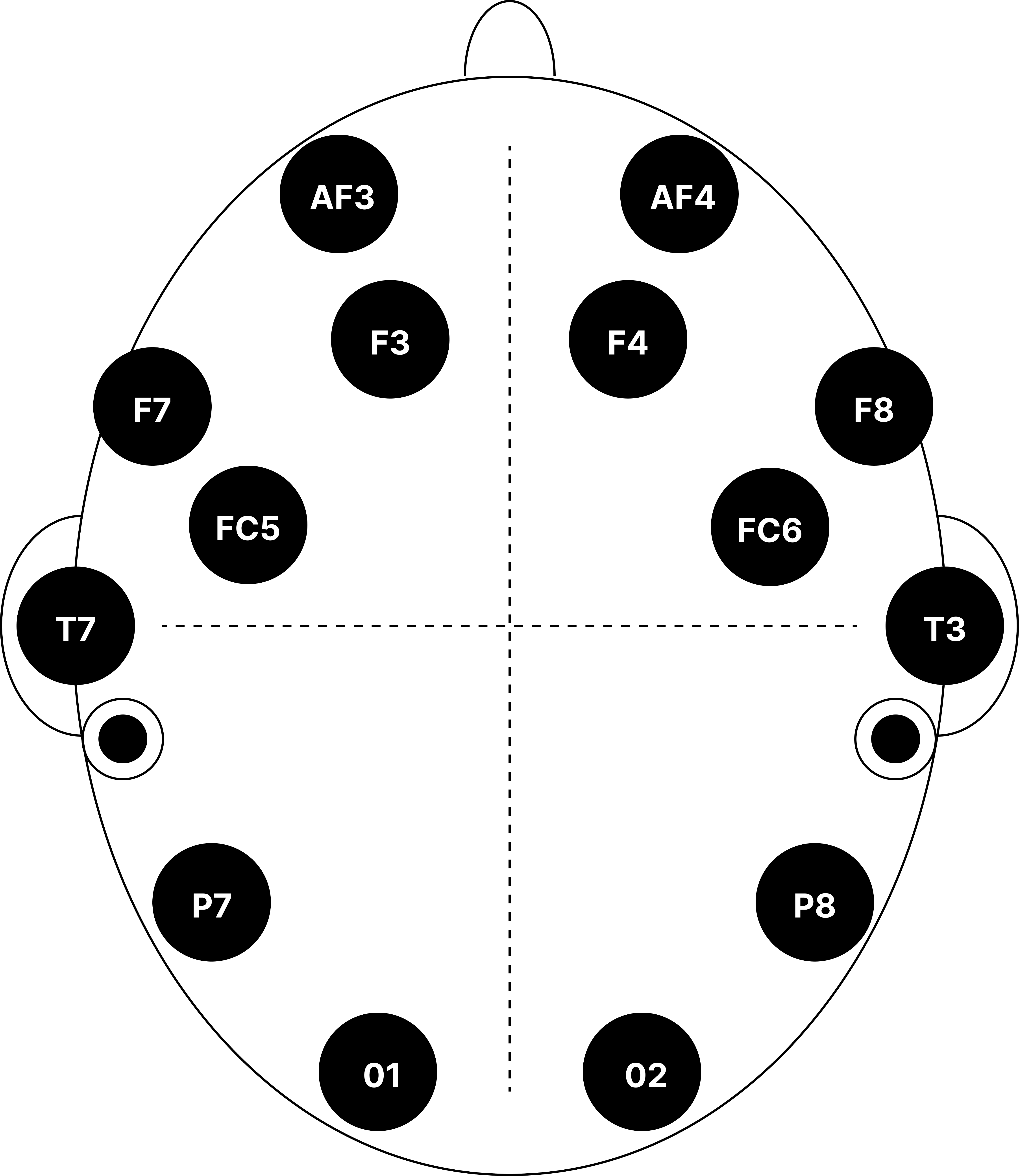}
\centering
\caption{EEG electrodes placement on head~\cite{jayarathne2020person}.} \label{fig1}
\end{figure}

The integration of EEG-based BCIs technology into the control of mobile robots, has led to a multidisciplinary approach which encompasses several fields, including neuroscience, computing, signal processing, and pattern recognition.
EEG signals are created by the activity of neurons in the
brain~\cite{nisar2013brain}.
The pattern of the EEG signals, correspond to the thoughts, emotions and
behavior of an individual~\cite{lam1997application,nisar2013brain}.
This technique is used to gauge variations in voltage attributable to the ionic current conduction within the neural networks of the cerebral cortex~\cite{mirza2015mind}.
Human brain can produced five major brain waves classified by frequency as illustrated in Section~\ref{sec:related}, Figure~\ref{fig1}.

The remainder of the paper is organized as follows: Section \ref{sec:related} discusses related works and summarizes relevant research in EEG technology and signal pattern recognition.
The needs for universal accessibility, ethical considerations for physical impairment individuals are described in Section~\ref{sec:discussion}.
Our design approach for physical impairments is introduced in Section~\ref{sec:design} and Section~\ref{sec:conclusion} concludes the paper.

\section{Related Works}
\label{sec:related}

Recent studies have been undertaken by researchers in the realm of BCIs , focusing especially on its usage in controlling wheelchairs~\cite{al2018review,antoniou2021eeg,kim2023literature,mirza2015mind,shahin2019wheelchair}.
These studies have demonstrated the feasibility and potential of using EEG- based BCIs to control wheelchairs, offering new possibilities for individuals with limited mobility.

\subsection{Wheelchair-BCI Control Systems}

Research in BCIs has opened up new possibilities for creating interactive systems that can translate human brainwaves into control signals for computer application devices~\cite{Lin2010EEGAE}.
This study~\cite{mirza2015mind} introduced a groundbreaking approach to mobility for individuals with disabilities by developing a mind-controlled wheelchair.
By harnessing the power of an EEG headset coupled with an Arduino microcontroller, the system captures brain and eye signals.
These signals are then translated into precise movement commands that the Arduino microcontroller executes, propelling the wheelchair.
Another study involved a double-blind randomized controlled trial where subjects use an EMOTIV Insight EEG headset to control a cursor on a computer screen.
The results indicate that allowing users to select their mental commands and training strategies enhances their control accuracy~\cite{siow2023human}.
In~\cite{fakhruzzaman2015eeg} EEG used to record brain activity, focusing on its application in motor imagery - the mental simulation of movement.
Nisar et al. reported that their system, which captures EEG signals using a 14-sensor headset and then transmits them to a robot through a computer interface, could control the robot's movements in real-time with a sensitivity and specificity exceeding 90\%~\cite{nisar2013brain}.

\subsection{EEG signal recognition}

Human brain can produce five major brain waves,
classified by their frequency ranges, known as Brain Rhythms.
These major waves range from low frequency (0.5 Hz) to high frequency (100 Hz).
These are known as delta(0--4 Hz), theta (4--8 Hz), alpha (8--13 Hz), beta(13--30 Hz), and gamma (30--100 Hz) waves~\cite{jayarathne2020person}.
Alpha waves are indicative of a relaxed, idle state of the brain.
Beta waves are observed when the brain is actively engaged in cognitive tasks such as thinking or problem-solving.
Theta waves are associated with stress, emotional tension, disappointment, deep meditation, or unconscious states.
Gamma waves are linked to consciousness and higher mental activity.
Lastly, Mu waves are typically observed in response to spontaneous motor activities~\cite{xavier2020exploratory}.
These various brain wave patterns have significant implications in the development of brain-computer interfaces, particularly in applications such as controlling wheelchairs, where different states or intentions like moving forward, turning, or stopping are crucial.

The EEG may show unusual electrical discharge when some abnormality occurs in the brain.
The measurement of placing the electrodes in the brain area, namely, frontal pole (Fp), frontal (F), parietal (P), temporal (T), and occipital (O), provides meaningful communication.
Even numbers and odd numbers as subscript have been decided to differentiate the brain's hemisphere.
The position of Fp2, F4, F8, C4, T4, T6, P4, and O2 electrodes indicates right hemisphere and Fp1, F3, F7, C3, T3, T5, P3, and O1 electrodes indicates left hemisphere, respectively.
The position of FZ, CZ, and PZ electrodes indicates the midline in frontal, central, and parietal regions~\cite{nanthini2017electroencephalogram}.

\subsection{Machine Learning Algorithms}

Advanced algorithms in signal processing and pattern recognition are used to interpret these brain signals accurately.
Machine learning algorithms, including supervised learning techniques such as Support Vector Machines (SVMs), neural networks, and deep learning models~\cite{essa2021brain}, are increasingly pivotal in the analysis of brain signals.
In supervised learning, algorithms are trained on labeled datasets to perform tasks like classifying mental states or detecting abnormalities in brain signals.
SVMs excel in classification tasks~\cite{iyortsuun2023review}, effectively differentiating between various mental states or identifying abnormal brain patterns, particularly in EEG signal analysis.
Neural networks, mimicking the human brain's structure, adapt flexibly for both classification and regression tasks in brain signal interpretation, such as predicting neurological disorder progression from brain imaging.
Deep learning models, especially Convolutional Neural Networks (CNNs), are adept at processing grid-like data, like MRI or fMRI scans, to recognize patterns indicative of brain conditions.
Recurrent Neural Networks (RNNs) and Long Short-Term Memory (LSTM) networks are suitable for sequential data, like time-series EEG or MEG signals, capturing the temporal dynamics of brain activity.
Additionally, autoencoders in unsupervised learning paradigms facilitate dimensionality reduction, crucial for visualizing high-dimensional brain data.
These algorithms' capability to process large datasets and discern intricate patterns significantly enhances our understanding of neural processes and augments the diagnosis and treatment of neurological disorders.
Limited studies have explored the combination of EEG-BCI data with the RF algorithm~\cite{antoniou2021eeg}.
In study~\cite{antoniou2021eeg} the experimental findings demonstrated that the RF algorithm surpassed other methods, achieving a high accuracy rate of 85.39\% in a 6-class classification scenario due to relatively fast training time, a small number of user-defined parameters and superior performance compared to the other five widely used classification algorithms.

Another research~\cite{badajena2023data} focused on enhancing smart wheelchair control through EEG and machine learning.
It proposed an attention meditation cost–benefit analysis (AMCBA) model, aiming to improve performance and decision-making in BCI systems~\cite{badajena2023data}.

\section{Discussion}
\label{sec:discussion}

Mobility constraints significantly impede the autonomy and overall well-being of individuals with physical disabilities.
The capacity for autonomous mobility is imperative for one's self-esteem and constitutes a pivotal component in the paradigm of aging in place~\cite{simpson2005smart}.
Wheelchairs emerge as indispensable aids for mobility~\cite{rushton2017understanding}.
However, manual wheelchair users still face challenges like limited mobility due to difficulty navigating uneven terrain, leading to restricted access to essential places and reliance on caregivers, which increases social isolation.
These issues, combined with physical demands and environmental barriers, underscore the need for improved infrastructure and supportive systems for better quality of life.
This limitation has given rise to the innovation of smart wheelchairs, equipped with robotics and human-computer interaction (HCI) capabilities, offering enhanced navigation for those with physical impairments~\cite{simpson2005smart}.

Our system is meticulously designed to pioneer an innovative paradigm in wheelchair mobility within the realm of digital twin environments, affording users the unprecedented ability to exert control through cognitive processes.
We conducted a thorough evaluation of the system's precision and responsiveness, employing EEG-based control mechanisms juxtaposed against conventional wheelchair navigation methods.
This comparison not only accentuates the system's enhancements in fostering user autonomy but also delineates its potential constraints.

\begin{table}[ht]

\caption{Comparison of EEG-Based control and conventional wheelchair navigation.}
\label{table:comparison}
\begin{tabular}{|p{0.2\textwidth}|p{0.35\textwidth}|p{0.40\textwidth}|}
\hline
\textbf{Aspect} & \textbf{EEG-Based Control} & \textbf{Conventional Navigation} \\ \hline
Precision      & High precision in capturing brain signals & Limited by manual control \\ \hline
Responsiveness & Immediate response to user intent & Delayed mechanical response \\ \hline
User Autonomy  & Enhanced by intuitive control & Restricted to physical ability \\ \hline
Constraints & Requires calibration and learning curve & Limited by physical constraints \\ \hline
\end{tabular}
\end{table}

In our exploration, we confront and dissect technical hurdles, including signal interference, and the pivotal role of machine learning algorithms in deciphering neural signals.
The incorporation of user feedback stands as a cornerstone of our research, offering invaluable insights into the system's practicality and ergonomic considerations, thereby paving the way for design optimizations.
Moreover, we engage in a critical discourse on ethical implications, with a special focus on privacy and data security concerns, underscoring the imperative to protect sensitive user information.
This ethical scrutiny extends to ensuring the responsible deployment and usage of this technology.
Looking ahead, we outline prospective research avenues aimed at refining the system's accuracy, enhancing its user interface, and ensuring its adaptability to accommodate a broad spectrum of disabilities.
We also ponder the broader societal ramifications of our work, advocating for the democratization of access to this cutting-edge technology.
Our discourse culminates in a contemplation of the transformative potential of such innovations, highlighting the necessity to bridge the gap between technological advancement and its equitable distribution among diverse user demographics, thereby magnifying its beneficial impact on society.

\subsection{Ethical Considerations}

The advancement of assistive technology brings with it a range of challenges and ethical dilemmas that must be carefully navigated.
A key ethical issue is finding the right balance between fostering independence and creating dependency in users of technology-assisted mobility devices.
The transition towards inclusive security should be comprehended as part of the wider evolution of security and privacy technology that is accessible to all~\cite{jones2023privacy}.

To effectively address the safety, privacy, and integrity concerns associated with BCIs in wheelchair technology, a comprehensive strategy is essential.
The risk of malicious interference with BCIs signals is significant, posing a threat to user safety.
It is crucial to ensure that BCIs not only accurately interpret user intentions but also reliably respond in complex environments.
The open transmission and processing of brain signals introduce vulnerabilities to unauthorized access and manipulation.
Safeguarding the integrity and confidentiality of these signals is critical to prevent harm.
This is particularly vital as inaccuracies or misinterpretations in signal processing could lead to dangerous situations, compromising the wheelchair's role as a safe mobility tool.

Moreover, the deeply personal nature of brain signals raises significant privacy issues.
It is imperative to protect these data from unauthorized access and ensure their use is strictly confined to intended mobility functions.
Addressing these multifaceted challenges demands robust encryption, strict safety protocols, and comprehensive privacy policies.
A collaborative approach involving researchers, technologists, and regulatory bodies is needed to develop standards and guidelines for the safe and ethical application of BCIs in wheelchair technology.

\subsection{Universal Accessibility}

Universal accessibility for individuals with physical impairments is a critical aspect of societal inclusion and well-being.
The concept of universal design, which aims to create environments and products that are accessible to all individuals, including those with disabilities, has gained significant attention in research and policy~\cite{iwarsson2003accessibility}.
This approach emphasizes the importance of integrating older and disabled individuals into mainstream society, highlighting the need for inclusive design strategies~\cite{dianat2018review} as well as human-centered design approach~\cite{morshedzadeh2022tapping}.
Furthermore, the significance of universal design in addressing the needs of individuals with physical impairments is underscored by the increasing recognition of the impacts of mobility impairments on transportation systems and the importance of designing inclusive transport networks~\cite{schmocker2009access}.
The need for standardized and readily accessible datasets to facilitate the management and comparison of global studies is emphasized in the context of metadata and taxonomic identification, reflecting the broader importance of standardized approaches in promoting universal accessibility~\cite{tedersoo2015standardizing}.

Given the importance of universal accessibility, it is notable to consider that people with physical impairments may also experience other conditions concurrently, such as visual, auditory, and cognitive impairments.
Recognizing and considering this intersectionality of disabilities is crucial for developers and researchers to offer the required assistance to individuals facing such conditions is essential.
It enables them to adopt a more inclusive and comprehensive approach in their work, thereby ensuring that the solutions and technologies they develop cater effectively to a broader spectrum of needs and challenges faced by individuals with disabilities.
In this context, the importance of developers' roles in advancing universal accessibility has gained increased prominence.

\begin{figure}
\includegraphics[width=\textwidth]{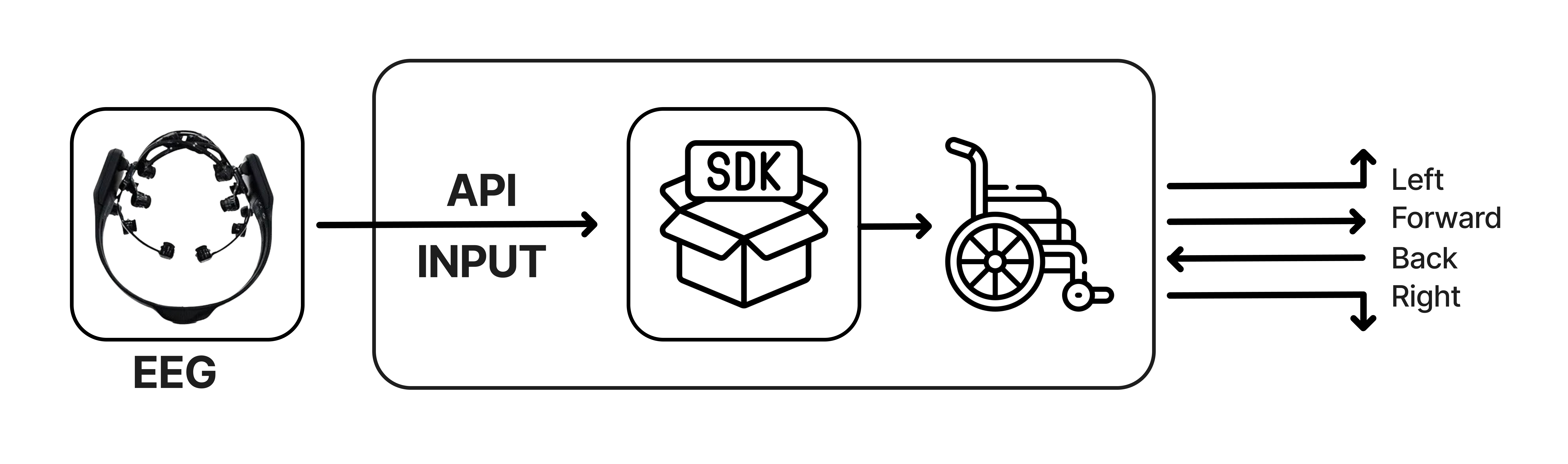}
\caption{A pipeline of the proposed design system.}
\label{fig2}
\end{figure}

\section{Design}
\label{sec:design}

We have developed a design system that illustrates the feasibility of controlling wheelchair navigation via neural input, specifically harnessing the user's brain activity.
Our approach facilitates a direct communication pathway between the brain and the wheelchair, bypassing the need for physical movement, which can be a significant advancement for those with severe mobility challenges.
In addition, our research incorporates a digital twin approach using the Unity Engine to simulate a 3D wheelchair.
This simulation can execute commands in real-time, such as moving forward or backward, directed by cognitive actions like ``Push'' and ``Pull'' that are identified from the user's brain activity.

This methodology is systematically delineated in Figure~\ref{fig2}, which provides a visual representation of the entire pipeline.
The process is executed through the following sequential stages:
\begin{enumerate}
\item
The initial stage involves capturing brain activity through an EEG headset, specifically the Emotiv EPOC+.
This device is equipped with 14 electrodes that are strategically placed on the user's scalp.
These electrodes detect electrical signals generated by brain activity, which are crucial for interpreting the user's intent regarding wheelchair navigation.
\item
Once the EEG signals are acquired, they are processed using EmotivPro and EmotivPro Analyzer, cloud-based platforms designed for analyzing brainwaves.
This step is vital for filtering out noise and extracting meaningful data from the raw EEG signals.
The platforms provide tools for visualizing and understanding the brain's activity patterns, which are instrumental in developing algorithms for interpreting navigational commands.
\item
The processed signals are then manipulated using the Cortex software development (SDK) provided by Emotiv.
This SDK is a cornerstone for integrating mental commands into the system.
It allows developers to access both raw and processed EEG data, facilitating the translation of brain activity into specific commands for wheelchair navigation.
The SDK's ability to handle real-time data processing is key to creating responsive applications that adapt to the user's mental commands instantly.
\item
The final step involves the Emotiv Unity Plugin, which is used to control a three-dimensional simulation of a wheelchair within the Unity environment.
This plugin acts as a bridge between the EEG data processing software and the Unity engine, enabling the real-time control of virtual objects—or, by extension, a physical wheelchair—through brain activity.
\end{enumerate}

\subsection{Pilot Study and Training Session}
Within the Emotive BCI visualizer, a 3D cube animation appears on the control panel.
Initially, training with neutral session which ensures readiness for the first command.
This phase is crucial for calibrating the system to recognize unique brain patterns associated with each command.
Successful training guarantees accurate control of the cube (or simulated wheelchair) per user intentions.
Then, four movements are programmed: pull (to move forward), push (to move backward), left (to turn left), and right (to turn right), as depicted in Table~ref{table:commands}.

\begin{table}[h]
\centering
\caption{Mental commands in a training session.}
\label{table:commands}
\begin{tabular}{|l|l|}
\hline
Command's Name & 3D Cube Movements \\ \hline
Push         & Move Forward \\
Pull         & Move Backward \\
Right         & Turn Right 90\degree \\
Left       & Turn Left $-90$\degree \\ \hline
\end{tabular}
\end{table}

\subsection{Digital Twins in Unity Game Engine}

Digital twins are virtual representations of physical objects or systems, and they have gained significant attention in various fields such as robotics, urban planning, manufacturing, and smart cities~\cite{dembski2020urban}.
These digital replicas are created using advanced technologies such as Unity, a popular game engine, which allows for the development of virtual environments and simulations~\cite{mignan2022digital}.
The use of Unity in digital twin development enables the training and visualization of trajectories for robots, the creation of virtual replicas of manufacturing cells, and the modeling of smart cities for urban planning and decision support~\cite{mignan2022digital}.
Furthermore, digital twins are not limited to static representations but have evolved to become actionable and experimentable, allowing for dynamic interactions and simulations~\cite{perez2020digital}.

\begin{figure}
\centering
\includegraphics[width=0.6\textwidth]{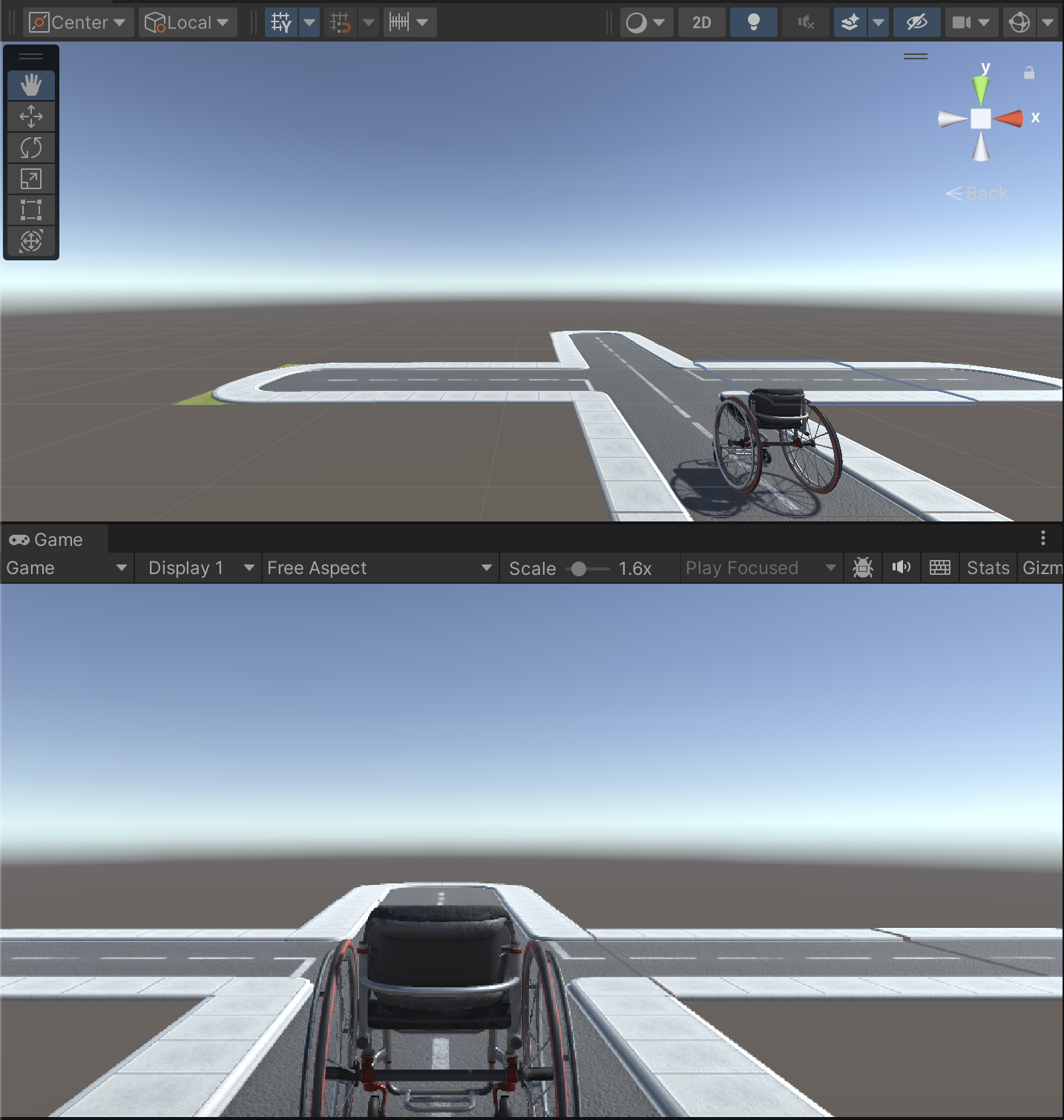}
\caption{A digital twin of Wheelchair 3D prototype in Unity engine.} \label{fig3}
\end{figure}

Based on the digital twin concept, we developed a 3D wheelchair simulation in Unity, which its built-in physics engine is employed to simulate real-world physics, thereby ensuring realistic interaction of the digital twin with its virtual environment (Figure~\ref{fig3}).
Additionally, custom C\# scripts are developed within Unity to replicate the operational behaviors of the entity, including movements, and interactions with environmental factors.
Inside the script, we initialized the Emotiv API and set up a connection to the headset.
We implemented functionality to process real-time data from the headset, focusing on mental commands that we want to use for controlling the wheelchair (e.g., move forward, turn).

Finally, we translated recognized mental commands into movements of the 3D wheelchair model.
This could involve changing the position and rotation of the model based on the commands detected.
Beyond real-time interaction, we can link Unity with the BCI Visualizer for training purposes and facilitate offline control by assigning mental commands from the cube to a 3D model of a wheelchair.

\section{Conclusion \& Future Works}
\label{sec:conclusion}

We proposed an innovative approach to wheelchair navigation by leveraging neural input from EEG signals.
This method represents a significant advancement in assistive technology, particularly for individuals with mobility impairments.
It showcases a promising step towards more independent and adaptable mobility solutions.
The research highlights the effectiveness of EEG-based systems in real-time control and underscores the importance of user-centred design in assistive technology.
While challenges remain, such as improving system responsiveness and broadening accessibility, this work lays a foundation for future innovations in the field, ultimately contributing to enhanced quality of life for those with physical disabilities.
The application of BCIs technology in this context holds the promise of revolutionizing standard wheelchair models, transforming them into more intuitive and responsive tools that are in sync with the cognitive commands of the user.

Future work for the EEG-Cortex SDK wheelchair control study should focus on enhancing EEG signal interpretation using advanced machine learning algorithms, improving user training for better system adaptability, and expanding the system's applicability to a broader range of physical disabilities.
Additionally, conducting long-term real-world evaluations to assess effectiveness and user satisfaction, and addressing scalability and cost concerns to increase accessibility, are crucial steps forward.

\begin{credits}
\subsubsection{\ackname} The authors express their sincere appreciation for the significant contribution made by Conor McGovern and  Caroline Marie Lassalle to the development of the prototype.

\subsubsection{\discintname}
The authors declare that they have no competing interests relevant to the content of this article.
\end{credits}

%
%
%
\bibliographystyle{splncs04}
\bibliography{Bibliography}
\end{document}